\documentclass{aa}
\usepackage{natbib}
\usepackage{txfonts}
\usepackage{graphicx}
\bibpunct{(}{)}{;}{a}{}{,}    
\newcommand{\Al}{$^{26}$Al\ }
\newcommand{\about}{$\simeq$}
\newcommand{\Msol}{M$_\odot$\ }
\newcommand{\flux}{ph~cm$^{-2}$s$^{-1}$\ }
\newcommand{\fluxrad}{ph~cm$^{-2}$s$^{-1}$rad$^{-1}$\ }
\begin{document}
\title{\Al in the inner Galaxy}

\subtitle{Large-scale spectral characteristics derived with SPI/INTEGRAL}

\author{R. Diehl \inst{1}  
 \and
 H. Halloin \inst{1}
 \and
 K. Kretschmer \inst{1}
 \and 
 A.W.Strong \inst{1}
 \and
 W. Wang \inst{1}
 \and
 P. Jean \inst{2}
 \and
 G.G. Lichti \inst{1}
  \and 
 J. Kn\"odlseder \inst{2}
 \and
 J.-P. Roques \inst{2}
  \and
  S. Schanne \inst{3}
  \and
  V. Sch\"onfelder \inst{1}
 \and
 A.von Kienlin \inst{1}
 \and 
 G. Weidenspointner \inst{2}
 \and 
 C. Winkler \inst{4}
 \and
 C. Wunderer \inst{5}
}

\institute{
Max-Planck-Institut f\"ur extraterrestrische Physik,
              D-85741 Garching, Germany 
 \and
 Centre d'Etude Spatiale des Rayonnements and Universit\'e Paul Sabatier, 31028 Toulouse, France
 \and
 DSM/DAPNIA/Service d'Astrophysique, CEA Saclay, 91191 Gif-Sur-Yvette, France	
 \and
  ESA/ESTEC, Science Operations and Data Systems Division (SCI-SD)  
  2201 AZ Noordwijk, The Netherlands 		  
 \and
  Space Sciences Lab., Berkeley, USA 		  
}%

\offprints{R. Diehl}

\mail{rod@mpe.mpg.de}
\date{Received 01 Oct 2005; accepted 17 Nov 2005}
\authorrunning{Diehl et al.}
\titlerunning{\Al in the inner Galaxy}%

\abstract
{We performed a spectroscopic study of the 1809~keV gamma-ray line
from $^{26}$Al decay in the Galaxy using the 
SPI imaging spectrometer with its high-resolution Ge detector 
camera on the INTEGRAL observatory.
We analyzed observations of the first two mission years, fitting spectra
 from all 7130 telescope pointings in narrow energy bins to 
 models of instrumental background and the \Al sky. 
 Instrumental background is estimated from independent tracers of cosmic-ray activation.
 The shape of the
 \Al signal is compared to the instrumental response to
 extract the width of the celestial line.We detect the $^{26}$Al line at \about~16~$\sigma$ significance.
The line is  broadened only slightly, if at all; 
we constrain the width to be below 2.8~keV (FWHM, 2$\sigma$). The average Doppler
velocities of \Al at the time of its decay in the interstellar 
medium ($\tau\sim$1.04~My) therefore are probably around 100~km~s$^{-1}$,  
 in agreement with expectations from Galactic rotation and
interstellar turbulence.
The flux (3.3~($\pm$0.4) 10$^{-4}$ph~cm$^{-2}$s$^{-1}$rad$^{-1}$) 
and spatial distribution of the
emission are found consistent with previous observations. 
The derived amount of \Al in the Galaxy is 2.8~($\pm$0.8)~M$_\odot$.

}

\maketitle  

\section{Introduction}

The detailed measurement of 1808.65 keV emission from Galactic \Al is one 
of the design goals of the INTEGRAL mission \citep{wink03}. 
\Al gamma-rays were
discovered in 1982 by HEAO-C \citep{maho82}, and since 
then have been considered to be direct
proof of ongoing nucleosynthesis in the Galaxy. From the 9-year mission
of the Compton Gamma-Ray Observatory, in particular 
from the COMPTEL imaging Compton telescope,
all-sky imaging in the 1.8 MeV gamma-ray line from \Al had been obtained
\citep{dieh95,ober97,knoed_img99,plue01}. 
From these measurements, we learn that \Al emission extends all along the plane 
of the Galaxy; hence
\Al nucleosynthesis is a common Galactic phenomenon rather than local
to the solar system. The irregular structure of the emission,
alignments of emission maxima with spiral-arm tangent, and comparisons
with tracers of candidate \Al sources all have pointed to the conclusion that
massive stars dominate \Al nucleosynthesis \citep{chen95,pran96,knoe_mod99,knoed_phd99}. 

The high spectral resolution of Ge detectors of 3~keV (FWHM) at the
\Al line energy of 1808.65~keV is expected to reveal more information about
the sources and their location through Doppler broadenings and shifts, induced from
Galactic rotation and from dynamics of the ejected \Al as it propagates in the
interstellar medium around the sources. 
This should add important new astrophysical perspectives, beyond what could
be learned from the large imaging database of the COMPTEL sky survey.

Significant broadening of the \Al line has been found by the GRIS team 
from its Ge detector instrument through a balloon experiment \citep{naya96}.
The reported intrinsic \Al line width of 5.4~keV corresponds to velocities above 
450~km~s$^{-1}$ for decaying \Al isotopes, if interpreted
dynamically, from Doppler shifts.
Various attempts were undertaken to understand how such high velocities could be maintained
on time scales of Myrs \citep{chen97,stur99}. It was clear that only rather unusual 
circumstances such as large interstellar cavities, or high fraction of \Al being
deposited onto grains near the \Al sources, could offer an explanation. 
With recent  space-based gamma-ray
spectroscopy experiments, in particular from the 
RHESSI \citep{lin02} and INTEGRAL \citep{wink03} missions,
it became apparent that the GRIS result  
may not be valid \citep{smit03,smit04,dieh03,dieh04}. The
interstellar medium  surrounding sources of \Al may be less
extreme; \Al is considered an important probe of these astrophysical sites.

In this paper, we discuss our analysis of data from the SPI spectrometer on INTEGRAL
with respect to the intrinsic \Al line width and to the large-scale properties 
of \Al emission in the inner Galaxy,
exploiting data from the first two years of the mission.

\section{Observations, data and analysis}

\subsection{Data selections and standard processing}

\begin{figure}[ht] 
 \includegraphics[width=0.5\textwidth]{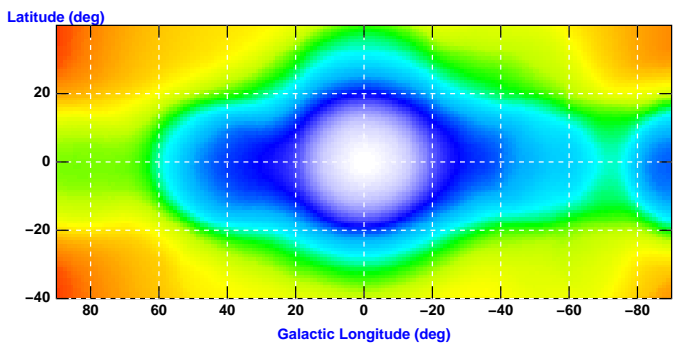}
 \caption{Exposure of the inner Galaxy for the data as selected from orbits 15-259
 for our study. At the Galactic center, the effective exposure is 4 Ms,
 at longitudes $\pm$50$^\circ$ 0.85 Ms (linear scale).
 }
 \label{exposure}
\end{figure} 

We analyze data from the SPI instrument, the coded-mask imaging spectrometer
instrument on INTEGRAL, with its 19-element Ge gamma-ray camera; 
the SPI instrument and its performance are described in \citep{vedr03,roqu03}. 
For the results reported here, we use data from the INTEGRAL Core Program
(which focuses on a survey of the inner Galaxy), 
our Open Program data, and data
that were publicly available at the time of this analysis. 
Our database spans INTEGRAL orbits 15--259, and our filtering to obtain
data free from contamination due to increased-background events or other anomalies
 yields a total observing time of 16.5~Ms from this set.
Filtering the data of glitches towards the perigee
near the radiation belts of the Earth, from solar-flare
periods, and from otherwise anomalous background conditions was
essential in order to obtain a sufficiently homogeneous database in terms
of background handling. 
We selected data within orbit phases 0.1--0.9 in order to exclude orbital phases near the
Earth's radiation belts, we excluded data where the saturated Ge count rate
exceeds 10$^5$~counts~s$^{-1}$ to reject anomalies, and we excluded 
solar-flare periods with a 1-day window triggered by the $>$30~MeV count rate of 
the GOES proton monitor exceeding a value of 0.5~Hz.

Our standard processing includes calibration of the energy scale once per orbit
by fitting Gaussians to strong instrumental lines. This attains a relative precision
of 0.05~keV at 1809~keV over these 1.5 years of data \citep{lonj04e}. The non-symmetric
line profiles, which develop from degradation of the Ge detectors due to 
cosmic-ray bombardement, result in shifts of the energy scale of 
a few tenths of keV at 
1809~keV. Due to calibration-line fitting with symmetric Gaussians, these detector
degradations mainly result in slight broadenings of the effective instrumental
line resolution; the offset in the absolute energy scale should be below 0.1~keV.
For each pointing of the instrument, standard processing assembles  the
energy spectra of each of the Ge detectors at 0.5~keV binning, 
together with detector deadtimes and various housekeeping data used for
data selections and for background modelling. 

In this study, we make use of
single-detector hits only, leaving coincident hits of more than one detector
aside. Although this reduces the overall detection efficiency by about 50~\%,
our background and spectral-response modeling is substantially more straightforward 
for this case.
Altogether our database covers the Galactic plane (pointings
within 30~degrees latitude) with 7130~pointings (135470~spectra), 
equivalent to a total deadtime-corrected exposure of 12.86~Ms, with, e.g., 
4~Ms at the Galactic center and 0.85~Ms at longitudes $\pm$50$^\circ$ 
(see Fig.\ref{exposure}).

\subsection{Background treatment}

Typically, source signals such as Galactic \Al contribute
only at the percent level, the remainder of the total signal being instrumental background.
Therefore the treatment of background is a key aspect of the achieved sensitivity and
suppression of systematics, as is the case for all gamma-ray instruments.
In principle, with SPI and its coded mask we can distinguish ``signal'' from 
``background'' by comparing the 19-detector amplitude patterns 
$ r_j=d_j/\sum_j d_j $
between sky observations and a suitable reference (j is detector index, d is counts);
instrumental background should be responsible for a pattern 
independent of instrument pointings. 
But diffuse sky emission such as from Galactic \Al is modulated 
rather weakly by the coded mask between 
successive pointings with typically 2$^\circ$ offsets, 
unlike for strong point sources. 
Therefore, the ``background reference'' pattern cannot be taken from 
pointings nearby in observing time, which is desirable to probe a similar background situation.  
For our large database, however, which spans a considerable range in time and sky-pointing directions, 
we can detect the celestial diffuse \Al signal when we use the time-averaged data for the 
background pattern reference (see Fig.~\ref{spec_onoff}). 
Independent data are preferable, e.g. when the instrument was pointed towards high latitudes,
or using detector count rates outside the energy range of interest, 
or using independent tracers of instrumental background.
Instrumental-background differences between such references
 and actual data may be expected, though; some time variation in the background model
should be allowed for by using additional parameters when fitting measurements. 

Following this general idea, we developed a background model for our study;   
more details will be described in a forthcoming paper (Halloin et al., in preparation).
We base our model on presumed tracers of instrumental background, and
optimize these to then establish the ``background'' detector ratio pattern per pointing.

First, we correlate a larger set of potential background tracers with the count-rate
variations in our energy band of interest around the \Al line. 
We choose the broad band of 1800--1820~keV for this
correlation, and the mission part where all 19 detectors of the camera were fully functional.
We sum over all 19 detectors to improve statistical precision 
for the detector data. This allows us to identify correlations of candidate background 
tracers with actual background variations in our camera with high significance. 
The best-correlated tracer is identified as the ``prime'' 
background tracer. Then we remove the information of this prime tracer
from all other candidate tracers through orthogonalization, and again correlate 
the (now additional to our prime tracer, or orthogonal-only) components 
of the other candidate tracers with our
count rates, to  iteratively find more ``next-best'' tracers. No more than 
three orthogonalized background-tracer components are needed for adequate background modelling
of the energy bands around the \Al line. 
This first step allows us to identify the necessary set of tracers and their decomposition order.
We find a hierarchy of (1) non-saturated Ge camera count rate, 
(2) cumulative saturated Ge counts since mission start,
and (3) plastic anticoincidence count rate.
In a second step, this decomposition scheme is applied to our spectroscopic dataset 
in 0.5~keV energy bins, separately for three time ranges to account for the two detector failures.
This determines for each tracer template the detector patterns of instrumental background 
as fitted to the measured data.
In this way, our background model is based on actual background-tracer data with their 
high statistical precision and high time resolution, yet the detector count ratios within 
the 19/18/17-element camera and per energy bin may be different.
In the final model fitting step (see next Section 2.3), only a few (3--10) background
parameters are adjusted to allow for gradual drifts of signal-to-background ratios
with time, and/or to account for normalization differences for the times where one or two detector
elements of our camera (detectors 2 and 17) failed.

\begin{figure}[ht] 
 \includegraphics[width=0.5\textwidth]{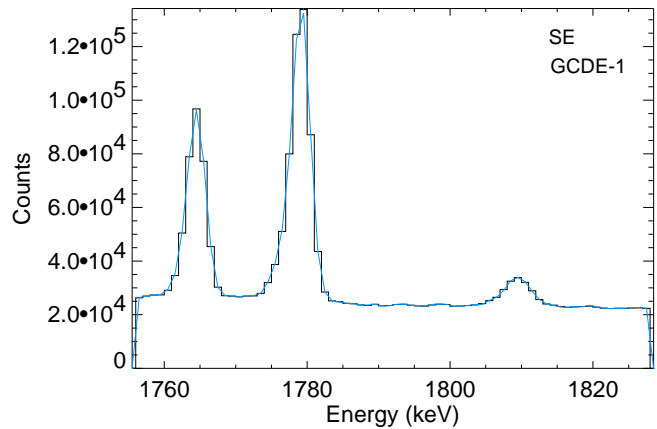}
 \caption{Raw data spectrum in the region around the \Al line, illustrating instrumental 
 line and continuum backgrounds (from single-detector events (SE) during the first
 inner-Galaxy survey (GCDE-1)). Line origins are radioactivities excited by
 cosmic ray bombardements, specifically $^{205}$Bi (1764.3~keV), $^{28}$Al (1779.0~keV),
 and $^{26}$Na (1808.7~keV) with $^{26}$Mn (1810.9~keV) for the composite feature underlying 
 the \Al line \citep{weid03}.
 Fitting these instrumental lines we typically find 
 centroids and widths of  (1764.4/3.1), (1779.0/3.2), and (1809.4/4.2), respectively.
 The intensity of the instrumental feature at 1809.4~keV is typically 12\% of the
 instrumental line at 1779~keV. 
 }
 \label{spec_raw}
\end{figure} 

\subsection{Spectra determination}

\begin{figure}[ht] 
 \includegraphics[width=0.5\textwidth]{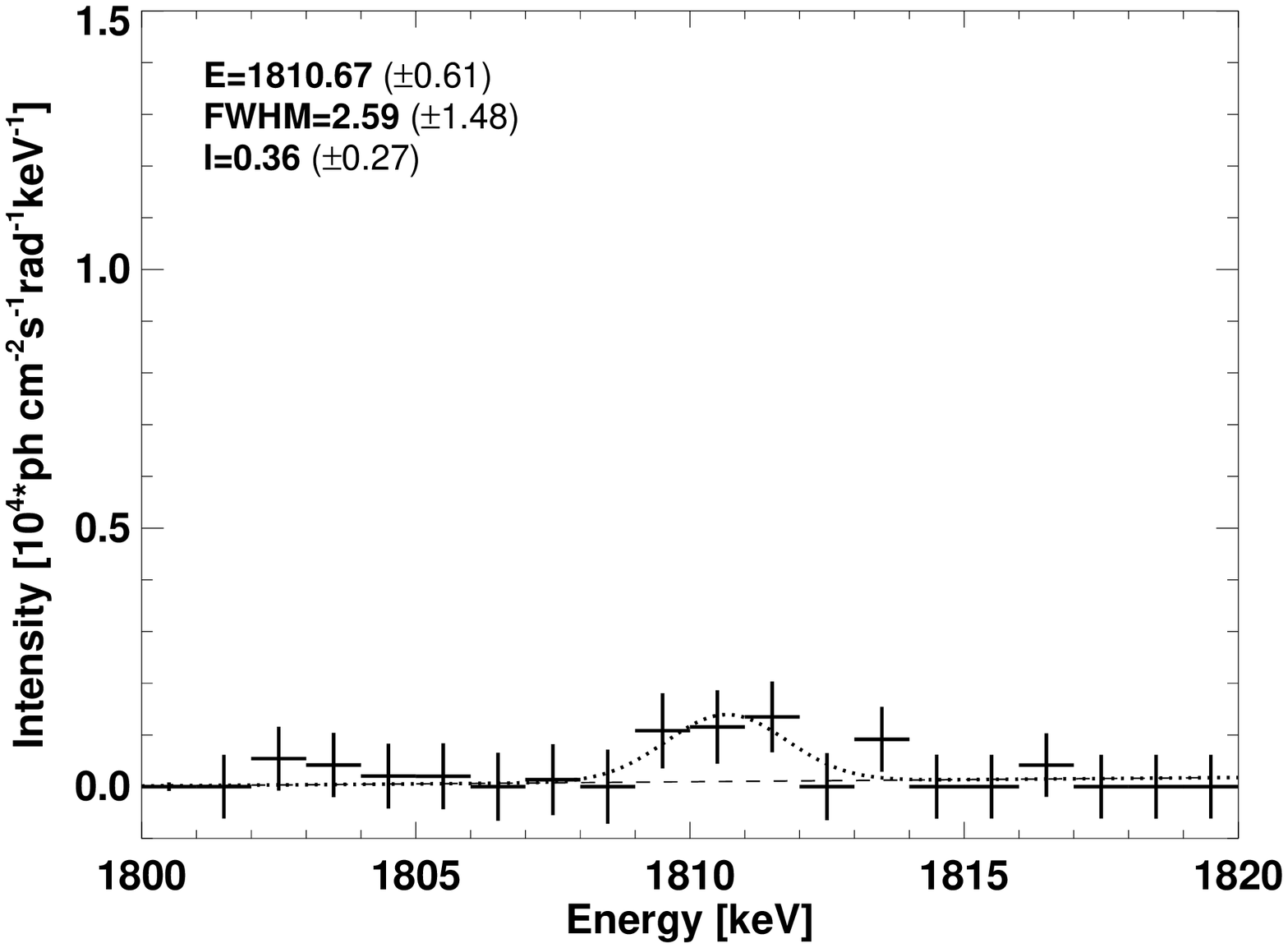}
 \includegraphics[width=0.5\textwidth]{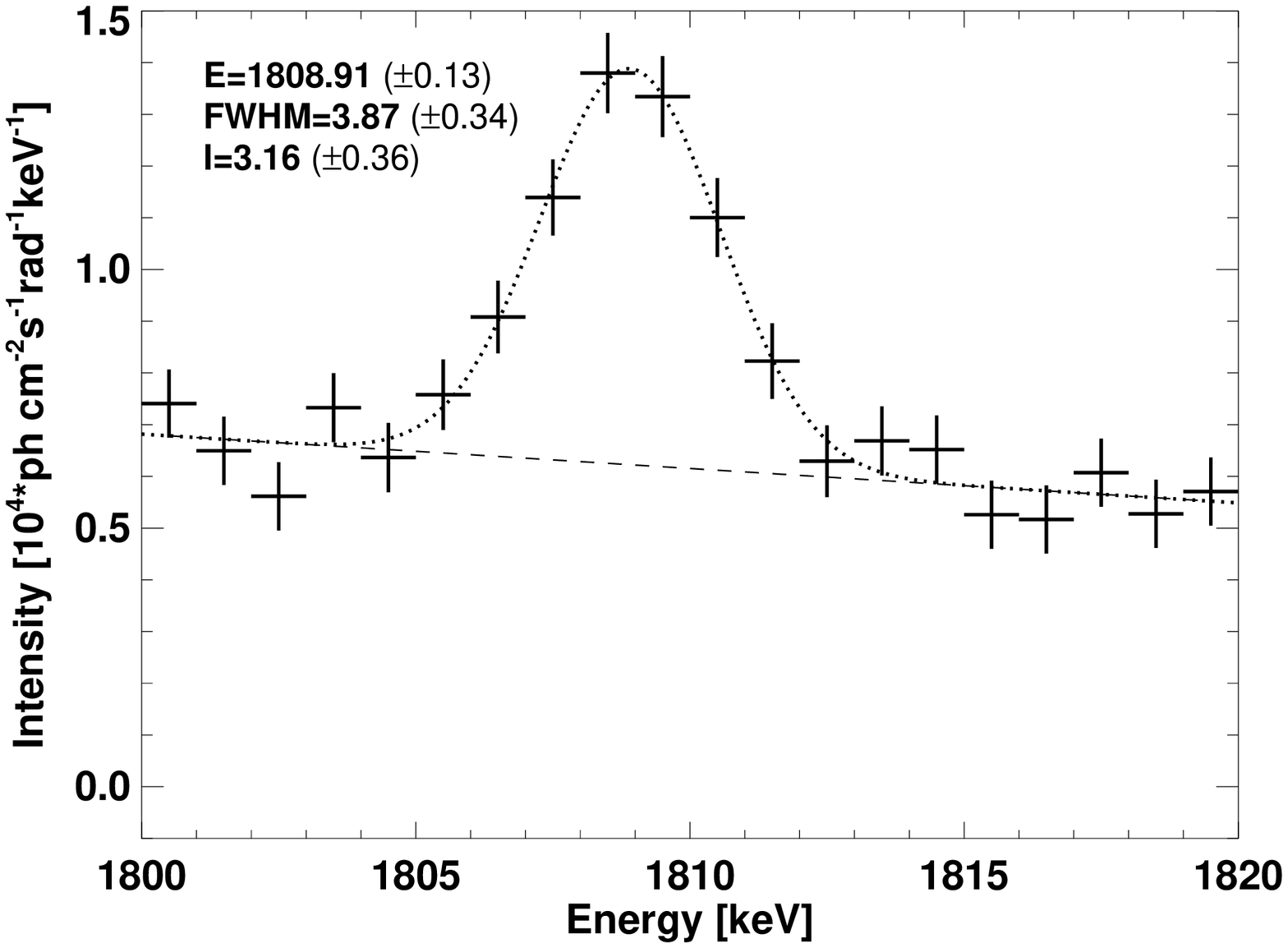}
 \caption{Test for spurious instrumental background leakage into spectra for celestial emission: 
 Sky model fitting was applied to high-latitude data (above), and to data from the inner Galaxy (below). 
 The sky intensity distribution was modeled from the COMPTEL \Al skymap, 
 and the pointing pattern of the inner Galaxy was adopted
 also for the high-latitude data. As a simple background model we
 use globally-averaged data;
 this background model still correlates with
 the measurements as taken along the Galactic plane, which causes
 non-zero intensities at all energies. Nevertheless, the celestial \Al line
 is revealed from the increased correlation with the skymap around
 1809~keV for the Galactic-plane data.
 }
 \label{spec_onoff}
\end{figure}

Our measured data consist of spectra per detector and pointing (the typical spectral signatures
are illustrated in the cumulative spectrum shown in Fig.\ref{spec_raw}).  
We derive our spectra of celestial emission by fitting our set of model components for background and 
a (set of) model (components) for the spatial distribution of celestial emission to our data, 
fitting amplitude coefficients independently for fine energy bins 
in the range 1800--1820~keV \citep[see also][ for method details]{stro05}.
The count data in our spectra per pointing and detector are modelled as:
$$d_{ijk}=\sum_{m,n}^{N_S}R_{ijk}^{mn}\sum_s\Theta_s S_s^{mn} + \sum_t\sum_b^{N_B}\Theta_{b,t} B_{b;ijk}$$
with {\it i,j,k} as indices for data space dimensions {\it energy, detector, pointing}, 
 {\it m,n} indices for sky dimensions {\it longitude, latitude},
  and $N_S$ sky model components $S$,   $N_B$ background model components $B$.
The skymaps $S$ are convolved into the data space of measured spectra for each
detector using observation attitude information and the instrumental response,
 which had been determined from Monte Carlo simulations and was adjusted 
 to prelaunch calibration measurements \citep{stur03}.
We assess the quality of the fit through checks on its global precision ($\chi^2$)
and on the residuals of the fit to our 135470 measured spectra.
Typical $\chi^2$ values per degree-of-freedom range from 1.5 in the instrumental
background line at 1779~keV  to 0.8--1.2 in the \Al line region. 
Typical statistial errors in fitted flux values per bin
 are 0.12~ph~cm$^{-2}$s$^{-1}$rad$^{-1}$~keV$^{-1}$.
 Systematic uncertainties can be estimated by increasing errors per data point until  
 the fit quality criterion $\chi^2$ yields a value of 1.0; we derive systematic uncertainty to 
 be 40\% of the value for statistical uncertainties.
Coefficients $\Theta$ for the background intensity (time variation allowed) and for the skymap intensity 
(constant in time) are derived,
the latter ($\Theta_s$) comprising the resultant spectrum of the signal from the sky (e.g. Fig.~\ref{spec_onoff}).

The celestial origin of the observed feature  near 1809~keV has been verified through identical analysis on
a dataset from high-latitude observations (Fig.~\ref{spec_onoff}), where the signal disappears. 
Moreover, the instrumental feature at 1810~keV (see Fig.~\ref{spec_raw}, leaking also
into the high-latitude data spectrum in Fig.~\ref{spec_onoff} (top)) is clearly offset from
the observed signal, and is significantly broader, 
due to its origin as a composite of local radioactivities \citep{weid03}.

\begin{figure}[ht] 
 \includegraphics[width=0.5\textwidth]{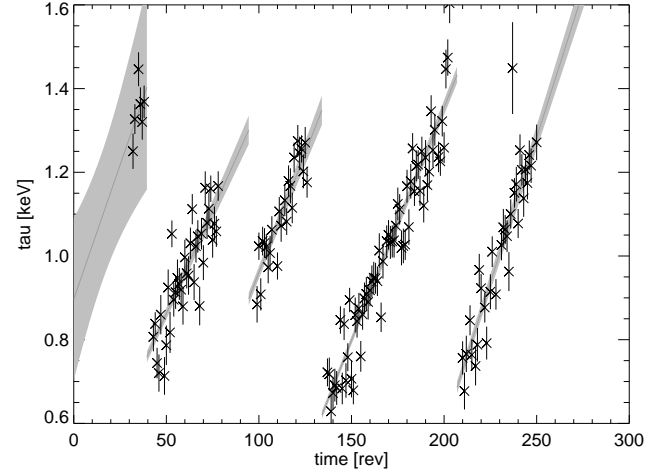}
 \caption{Evolution of the detector resolution with time for the region around 1800~keV.
 The width $\tau$ of the one-sided exponential which we use to describe degradation, 
 averaged over all 19 detectors, 
 evolves with time (expressed here in 3-day intervals of the satellite orbits) 
 due to degradation and detector annealings. 
 The grey areas represent the linear approximations which
 we adopted for modelling detector degradations, with their uncertainties (Kretschmer et al., in preparation).
 }
 \label{degradation}
\end{figure} 

The instrumental line-shape variation due to degradation of detectors was 
determined during the mission by fitting a specific spectral response to instrumental
lines which captures a degradation parameter $\tau$ (see Fig.~\ref{degradation}).
This spectral response function consists of a Gaussian shape which characterizes
each particular detector's intrinsic resolution, and a one-sided exponential
function extending from the peak of the Gaussian towards lower energies, 
which characterizes the pulse height losses due to detector degradation 
through its width $\tau$:
$$ s_i = s_0 \cdot \int_0^\infty \left( {{1}\over{\sqrt{2 \pi}\sigma}} \cdot 
   e^{-{{(E_i+E^\prime-E_0)^2} \over {2\sigma^2}}} 
    \cdot {{1}\over {\tau}}
    e^{-{{E^\prime} \over {\tau}}}\right) dE^\prime$$
with $s_i$ the amplitude per energy bin $E_i$, $E_0$ the photopeak line energy,
and $\sigma$ the instrumental resolution of the Ge detector. 
By fitting many calibration lines over the SPI energy range, the time-variable
degradation has been determined (see Fig.~\ref{degradation}).
In this analysis, we use this spectral-response behaviour to accumulate
the expected instrumental line shape, which we use in fitting the celestial
signal in our \Al line spectra and to derive constraints on the additional
broadening caused by intrinsic velocity variances of decaying \Al 
(e.g. Fig.~\ref{spec_width_cosmic}, see below). 
We plan to take this time-variable spectral response directly
into account in our instrumental response when convolving the input sky, and then
also by our modeling of instrumental features of the background  (Kretschmer et al., in preparation). 

In fitting non-analytical spectral shapes to our spectra, such as our time-integrated
instrumental-response line shape convolved with intrinsically-broadened \Al emission,
gradient-driven fit algorithms are inadequate - in particular when we aim
to determine a parameter like the intrinsic \Al width, which is convolved with the
instrumental line profile before being compared to our flux values per energy bin.
The probability distribution in this case is very asymmetric, but we wish to
perform quantitative statistical analysis of our findings. This is not possible
with tools that inherently assume symmetric and smooth probability distributions for
parameter fluctuations. We make use of the Monte Carlo Markov Chain method,
preprocessed by simulated annealing \citep{metr53,neal93}. 
This method migrates through the plausible
parameter space as guided by prior knowledge about the parameters, but in a
random-walk like fashion; for large numbers of samples, this assembles the
parameter value probability distributions rather well also for less idalized
cases (see Fig.~\ref{spec_width_cosmic} and next Section), allowing us to
derive an upper limit for the intrinsic \Al line width. Varying priors
within plausible ranges has an insignificant impact on our obtained numerical value.

\section{Results}

\begin{figure}[ht] 
 \includegraphics[width=0.5\textwidth,clip]{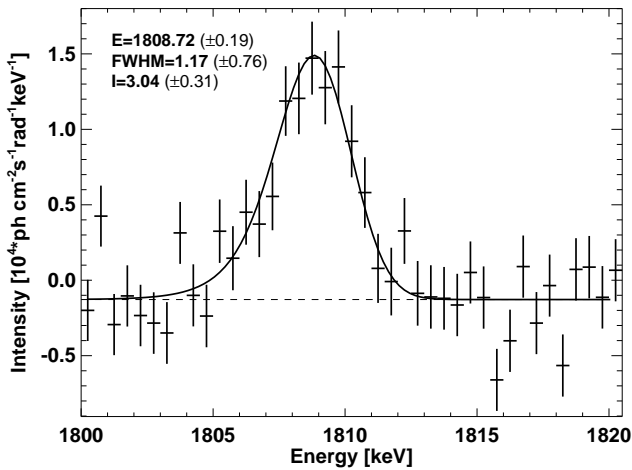}
 \caption{Spectrum derived from sky model fitting
 using COMPTEL \Al MaxEnt image and the ``orthogonalized tracers'' background model.
 The fit (solid line) uses the shape of the instrumental resolution as 
 it results from cosmic-ray degradation and annealings during the time of
 our measurement (see Fig.~\ref{degradation}), convolved with a Gaussian for the instrinsic \Al width,
 as fitted by MCMC. The \Al line width constraints are highly asymmetric
 (see Fig.~\ref{spec_width_cosmic}). Intensities are quoted in units of
 10$^{-4}$~ph~cm$^{-2}$s$^{-1}$rad$^{-1}$keV$^{-1}$.  
 }
 \label{spec_modfit_instrum}
\end{figure} 

By fitting our set of observations at 0.5~keV wide energy binning 
with the sky intensity distribution of \Al 
as imaged by COMPTEL together with our background model based on orthogonalized 
background tracers, we obtain the spectrum for
\Al emission from the inner Galaxy shown in Fig. \ref{spec_modfit_instrum}.
The \Al line is detected at 16$\sigma$ significance.
The line component above the linear component of our spectral fit determines
the intensity of observed $^{26}$Al. 
The total \Al gamma-ray flux as determined for the inner Galaxy region
($-30^\circ < l < 30^\circ$ and $-10^\circ < b < 10^\circ$, i.e. the flux
in the inner radian) obtained from this fit is 
3.3~($\pm$0.4)~10$^{-4}$~ph~cm$^{-2}$s$^{-1}$rad$^{-1}$. 
This is consistent with both the value of about 
4~10$^{-4}$~ph~cm$^{-2}$s$^{-1}$rad$^{-1}$ concluded
from previous measurements \citep{pran96} and in particular the COMPTEL imaging-analysis
value of 2.8~($\pm$0.4)~10$^{-4}$~ph~cm$^{-2}$s$^{-1}$rad$^{-1}$ \citep{plue01phd,ober97}. 
When we vary the model sky among plausible models, we obtain slightly 
different flux values, within expectations for the respective candidate \Al source maps.
The variance of flux values in our sample  (see Table \ref{table:1} and \citet{dieh05Nat})
is 1.2~10$^{-5}$~ph~cm$^{-2}$s$^{-1}$rad$^{-1}$ or 4\%;
we adopt this as our systematic flux uncertainty.

\begin{table}
\caption{\Al line results for different sky distribution models} 
\label{table:1} 
{\centering 
\begin{tabular}{c c c c c} 
\hline\hline 
 Sky Model &  d$\lambda$$^1$   &  I$^2$ &  dI$^2$ &  width$^3$  \\
\hline 
 exponential disk$^4$ &  n/a &  2.88 &  0.33 &  1.09 \\
 young disk$^5$ &  -0.62 &  3.22 &  0.43 &  1.19 \\
 free electrons$^6$ &  -1.49 &  3.54 &  0.36 &  1.05 \\
 Ne2001(180)$^7$ &  -2.87 &  3.22 &  0.38 &  1.34 \\
 Ne2001(140)$^8$ &  -2.22 &  3.06 &  0.41 &  1.22 \\
 dust 100$\mu$m$^9$ &  -1.14\ &  4.01 &  0.42 &  1.17 \\
 COMPTEL \Al$^{10}$ &  -0.70 &  3.04 &  0.31 &  1.17 \\
\hline 
\end{tabular}}
\\
$^1$ Fit quality: likelihood ratio difference to first model \\
$^2$ in units of 10$^{-4}$~ph~cm$^{-2}$s$^{-1}$rad$^{-1}$ \\
$^3$ mean of celestial line width (see Fig.\ref{spec_width_cosmic}) \\
$^4$ exponential disk, scale radius 4 kpc, scale height 180 pc \\
$^5$ young disk from \citet{robi03}, scale height 125 pc \\
$^6$ free electron spiral arm model from pulsar dispersion measurements \citep{tayl93}, scale height 180 pc \\
$^7$ free electron model from \citet{cord02}, no thick disk, scale height 180 pc \\
$^8$ as $^7$, but scale height 140 pc \\
$^9$ IRAS 100 $\mu$m, from {\it skyview} \citep{whee91}, after subtraction of Zodiacal light \\
$^{10}$ ME map from \citet{plue01}
\end{table}

The detection of the celestial \Al line is significant ($>3\sigma$)  
in 6 of the 0.5 keV-wide bins covering the \Al
line; this allows us to derive line shape details for the \Al line. 

First, the line centroid is determined at 1808.72 ($\pm$0.19(stat)$\pm$0.1(syst))~keV from fitting
our model shape to the observed feature (see above). This is
well within the laboratory value for the \Al line of 1808.65~(7)~keV \citep{ToI}. The variance of
line centroids for the different models of Table \ref{table:1} is 0.002~keV.

\begin{figure}[ht] 
 \includegraphics[width=0.5\textwidth,clip]{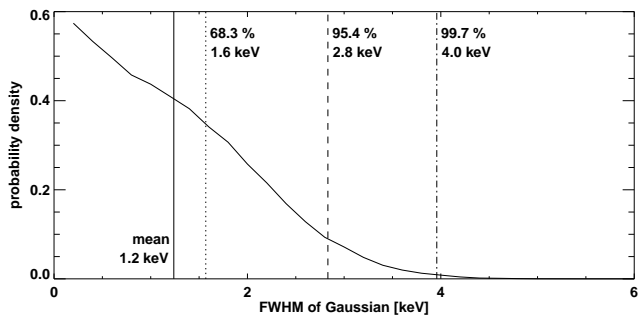}
 \caption{Probability distribution for the \Al intrinsic line width
 as fitted in Fig.~\ref{spec_modfit_instrum}. The formal results (Fig.~\ref{spec_modfit_instrum})
 derived from the mean and width of this distribution are inadequate descriptions
 of this highly asymmetric probability distribution. Rather, an upper limit
 of 2.8~keV (2$\sigma$) applies, and an intrisically even much narrower line is
 fully consistent with our measurement. 
 }
 \label{spec_width_cosmic}
\end{figure}

Second, the \Al line shape rather closely resembles the one expected from instrument properties
such as intrinsic resolution and degradations as experienced between annealings.
We fit the effective accumulated instrumental line shape, convolved with
a Gaussian for the celestial \Al line broadening, to our spectrum. The fitted 
parameters are the line centroid, the intrinsic width of celestial $^{26}$Al, 
the intensity of the line, and two parameters for the underlying continuum.
The intrinsic line width appears to be rather small. But this parameter
is not as ``well-behaved'' as the others: The parameter probability
distribution should be symmetric as for a Gaussian, its width then reflects 
the uncertainty of the fitted parameter value. The probability
distribution for the \Al line width, however, peaks at small values near 0.2~keV, with a minor reduction
towards zero intrinsic line width, and gradually decreases towards larger line widths
(Fig.~\ref{spec_width_cosmic}). Formal determination of the mean of this probability distribution
(to yield a ``fitted width value'') and its width (to yield the ``width uncertainty'')
is not appropriate. (Nevertheless, we show these results in spectra
Figures to at least coarsely describe the intrinsic line broadening). 
Therefore, we obtain the probability distribution  
from the Monte Carlo Markov Chain parameter fitting; we integrate this distribution 
over the desired probability fractions to obtain
upper limits for the intrinsic \Al line width, see Fig. \ref{spec_width_cosmic}: 
For a 95\% probability
equivalent to a ``2$\sigma$'' limit, we thus obtain a limit of 2.8~keV from our 
SPI measurement, while the GRIS value of 5.4~keV has a probability of $3~10^{-5}$.
It turns out that the
spectral details (line position and width) are practically identical 
for different adopted sky distributions, 
while \Al flux values vary by about 4\% among different plausible models \citep{dieh05Nat}.

\section{Discussion}

Our new measurement of the inner-Galaxy integrated \Al emission yields a flux
that translates into a Galactic steady-state mass of 
2.8 ($\pm$0.8)~\Msol \citep[see ][ for details and implications]{dieh05Nat}. From
a comparison to earlier measurements \citep[see][]{pran96}, our flux value is on the
low side of the average among experiments. This trend for imaging versus
large field-of-view experiments may suggest that either imaging helps to 
suppress instrumental background leakage, or that there is some large-scale
extended \Al emission from nearby sources. 

As a first step towards spatially resolved spectroscopy, we separately determined 
spectra for the inner region of the Galaxy (-40$^\circ<l<40^\circ$, -10$^\circ<b<10^\circ$)
using a smooth exponential-disk model with scaling parameters 4~kpc (radius) and
180~pc (latitude extent), which we split into three longitude segments at 
-10$^\circ$ and 10$^\circ$. This obtains
spectra for the different segments \citep[][ and referenecs therein]{dieh05Nat}. 
Fitting the \Al line as above, we obtained 
differences in line centroids compatible with expectations from Galactic rotation:
the \Al line is slightly blue-shifted at negative longitudes
and red-shifted at positive longitudes. This supports the view that the observed \Al emission arises from
the inner Galaxy region, rather than foreground. This reaffirms the above Galaxy-wide steady-state
interpretation of the measured \Al flux and determination of the total \Al mass in the Galaxy \citep{dieh05Nat}.
Furthermore, these spatially separated spectra also support the asymmetry of the
\Al emission in the inner Galaxy indicated in COMPTEL images \citep{plue01}, 
in that the 4$^{th}$ quadrant appears brighter than the 1$^{st}$.

It is evident that line broadenings of a few keV disagree with our data
from the inner Galaxy
(Figs. \ref{spec_modfit_instrum} and \ref{spec_width_cosmic}); 
thus, velocities of decaying \Al isotopes are
modest. The GRIS measurement of 5.4~keV line broadening appears inconsistent
with all other measurements to date and is clearly ruled out by our result.
 
A line broadening of 0.8~keV corresponds to thermal Doppler velocities of 100~km~s$^{-1}$, 
as a reference. In the inner
Galaxy, Galactic rotation alone leads to Doppler shifts, which are 
particularly pronounced towards longitudes $\pm$30$^\circ$, and range
up to 0.25~keV \citep{kret03}; a line broadening of about 1~keV had
been estimated if integrated over this inner region of the Galaxy \citep{kret03}. 
Are our line-shape constraints consistent with
Galactic-rotation effects and standard interstellar velocities?

\begin{figure}[ht] 
 \includegraphics[width=0.5\textwidth]{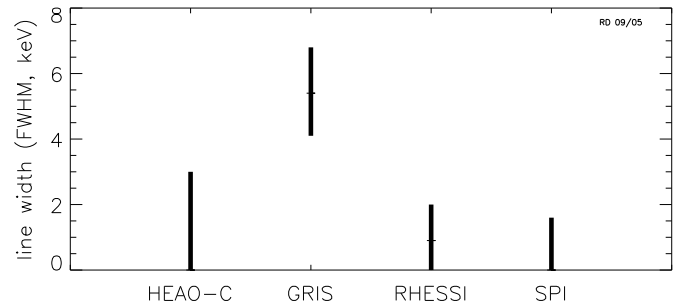}
 \caption{Constraints on the width of the \Al line from different instruments
 (1$\sigma$ error bars).
 }
 \label{linewidth_instruments}
\end{figure} 

We believe so \citep{dieh05Nat}, and consider the measured line width of \Al from the inner Galaxy to be
consistent with Galactic rotation and modest (see below) interstellar-medium turbulence
around the sources of $^{26}$Al.    
We thus confirm earlier results obtained by HEAO-C, RHESSI, and 
SPI on INTEGRAL \citep{maho82,smit03,dieh04}
 (see Fig.~\ref{linewidth_instruments}). 
This is reassuring: neither the suppression of deceleration of \Al before decay
in large, kpc-sized cavities around its sources nor other
exotic explanations   \citep{chen97} are required
to account for a large broadening of the \Al line. 

How much line broadening from ISM is plausible?
Some broadening is expected from the apparent existence of major
interstellar cavities in the regions of massive-star clusters \citep{oey96}, and from the 
ejection kinematics of these presumed \Al sources, i.e. winds and supernovae.
WR wind velocities are 
1200~km~s$^{-1}$ \citep{vink05} or higher \citep{plue01phd},
and models for ejection of \Al by core-collapse supernovae predict velocities 
in the same range \citep{hera94}. Therefore it will be interesting to test with SPI
on INTEGRAL whether we can observe such effects for localized regions of \Al
emission. Here sources may not yet have obtained a long-term equilibrium with
their surrounding ISM, and \Al may thus be preferentially decaying 
from its initial and fast phase, rather than already being slowed down to 
normal ISM velocities in the 10~km~s$^{-1}$ range. The first SPI results
of a modestly-broadened \Al line in the Cygnus region \citep{knoe04c},
as well as the COMPTEL result of the Orion region (where \Al emission appears offset from its sources
in the Orion OB1 association towards the Eridanus cavity \citep{dieh02}),
are hints that \Al streaming occurs at unusually high velocities and 
kinematics may be peculiar around regions with a rather young 
(few Myr) population of massive stars.
 
Our constraints on large-scale integrated broadening of the \Al line from the inner Galaxy can be
interpreted in terms of interstellar-medium characteristics, if we adopt the contribution
of Galactic rotation from a model: For 1~keV broadening due to Galactic rotation \citep{kret03}, 
the intrinsic width from ISM turbulence would be 1.2~keV to obtain our 1$\sigma$ line width
limit of 1.6~keV. This corresponds to 300~km~s$^{-1}$ for a 2$\sigma$ limit on ISM velocities,
leaving typical velocities up to \about~100~km~s$^{-1}$, well within the acceptable range. 
Therefore we conclude that, within uncertainties,
the average velocities of decaying \Al in the Galaxy are  
probably not in excess of typical values for the ISM near massive stars.

\begin{acknowledgements}
This paper is based on observations with INTEGRAL, an ESA project with instruments and science 
data centre funded by ESA member states (especially the PI countries: Denmark, France, 
Germany, Italy, Switzerland, Spain), Czech Republic and Poland, and with the participation 
of Russia and the USA. 
The SPI project has been completed under the responsibility and leadership of CNES/France.
The SPI anticoincidence system is supported by the German government through DLR grant 50.0G.9503.0.
We are grateful to ASI, CEA, CNES, DLR, ESA, INTA, NASA and OSTC for support.
\end{acknowledgements}

\bibliographystyle{aa}             

\end{document}